\documentclass[aps,preprint,amsmath,amssymb]{revtex4}
\usepackage{amsmath,amssymb}
\usepackage{graphicx}
\usepackage{booktabs}
\usepackage[usenames, dvipsnames]{color}
\usepackage{multirow}
\usepackage{setspace}
\usepackage{rotating}
\usepackage{afterpage}
\usepackage[ pdftex, plainpages = false, pdfpagelabels, 
                 pdfpagelayout = useoutlines,
                 bookmarks,
                 bookmarksopen = true,
                 bookmarksnumbered = true,
                 breaklinks = true,
                 linktocpage=true,
                 pagebackref=false,
                 colorlinks = true,
                 linkcolor = BrickRed,
                 urlcolor  = blue,
                 citecolor = BrickRed,
                 anchorcolor = green,
                 hyperindex = true,
                 hyperfigures
                 ]{hyperref}

\begin{document}

\title{Opportunities in Delivery of Preventive Services in Retail Settings}

\author{\href{http://necsi.edu/faculty/bar-yam.html}{Yaneer Bar-\!Yam}, Dion Harmon, Keith Nesbitt, May Lim, Suzanne Smith}
\affiliation{\href{http://www.necsi.edu}{New England Complex Systems Institute} \\
238 Main St.~Suite 319 Cambridge MA 02142, USA }
\author{and \\ Bradley A. Perkins}
\affiliation{Centers for Disease Control and Prevention \\
1600 Clifton Rd. Atlanta, GA 30333}

\date{August 3, 2008; released June 28, 2012} 

\begin{abstract}
Recommended clinical preventive services are not being delivered despite well-documented benefits. Here we show that transferring simple and repetitive preventive services to nurse-staffed retail clinics provides an opportunity for dramatically improving their delivery.  For each of 35 high-benefit, cost-effective preventive services, we identify required training, number of repetitions, and time and cost for full coverage in the US.  We determine that full delivery through physician-based practices would require an unrealistic 400,000 full-time personnel. We estimate the efficiency gains from implementation at nurse-staffed clinics at retail locations for 28 services. Widespread adoption would result in a five-fold reduction in variable costs and three-fold reduction in personnel. By elevating the benefit-to-cost ratio, retail implementation can expedite widespread prevention coverage and help transform US healthcare. 
\end{abstract}

\maketitle

\section{Introduction}
\label{sec:intro}
Improving the US healthcare system requires not only providing care to the uninsured but also addressing a set of linked organizational and motivational issues, including enhancing the role of wellness and prevention. It is widely acknowledged that system organization is key to healthcare improvement [1]. Previous studies [2, 3] have pointed to separating wellness and prevention services from acute care as a central component of healthcare transformation. Even while chronic and preventable diseases have become the dominant cause of loss of life, incentives and perceptions limit the delivery of services that could prevent them [2]. A complex systems analysis implies that the optimal organizational structure depends critically on the scale of repetition and complexity of tasks to be performed. Applied to healthcare, the separation of acute care from selected preventive services should enable improved matching of organizational structure to function [3--5].

The distinction between simple repetitive and complex tasks is apparent in the proper and improper application of efficiency. Complex tasks---including most medical diagnosis and treatment---require extensive training and careful decision making to determine which one of many possible actions should be performed. Efficiency is detrimental to complex tasks, as shorter times and streamlining through standardization curtails the necessary decision-making. In contrast, simple repetitive tasks are amenable to rapid execution by streamlined processes. Without such efficiency the necessary repetitions may not be achieved due to insufficient manpower and other resources. Conserving limited resources in simple repetitive tasks enables those resources to be utilized for high complexity tasks. The distinction between simple and complex is manifest in the allocation of tasks within a hospital ranging from laundry to diagnosis. It is also apparent in a great success of public health---smallpox eradication---that relied upon two processes: mass immunizations and ``surveillance and control'' [6]. The former, a simple repetitive task, was made highly efficient to enable 100 million vaccinations over five years. The latter, a high-complexity task, was performed by specially trained teams to identify individual cases and vaccinate their households and close contacts.

A focus of public health prevention efforts is achieving higher levels of delivery of clinical preventive services [7--9]. 100,000 additional lives could be saved annually by increasing five services to 90\% delivery [10]: tobacco cessation counseling, aspirin chemoprophylaxis, influenza immunizations, colorectal cancer screening, and breast cancer screening. However, these long standing recommendations [11] have not achieved their full potential impact and therefore we consider the obstacles to and opportunities for improvement based upon an analysis of organizational structure.

Retail clinics have been introduced in the US [12] as a response to the need for convenient and affordable care, and are gaining popularity as a mechanism of simple care delivery. Retail clinics, also known as convenient care clinics, are medical institutions established within a retail setting, such as a shopping mall. They are generally staffed by nurse practitioners or physician's assistants---individuals who have medical training and are able to write prescriptions. Distinct from ``urgent care'' centers, they follow a retail model---they provide only relatively simple services that can be delivered rapidly by non-physician providers, and allow payment by insurance as well as directly by consumers [13]. Retail clinics diagnose and treat a variety of common ailments: the common cold, the flu, ear infections, allergies, injuries, rashes, etc. They also provide some preventive services. They may be open for extended hours and seven days a week. Retail clinics join pharmacies and opticians in using retail locations for a medical purpose.

As of 2011 there were over 1,200 retail clinics in the US [14]. Some are independently operated, while others are managed as part of pharmacy or retail chains, including CVS, Walgreens, and Target, within their store locations. It is estimated that 1 in 6 Americans have already visited a retail clinic, while almost half are receptive to the idea; young, healthy individuals are particularly receptive to the use of retail clinics [15]. The accepted purpose of retail clinics is to provide more convenient locations and times, and lower prices than traditional physician practices. 

As an innovation in the healthcare system, retail clinics have provoked concerns reflected in prominent recommendations for investigation and legislation [16]. The debate has not been informed by a framework adequate to analyze the potential contributions of various service delivery models. In this chapter we propose that the low-complexity, large-scale nature of a retail clinic's practice allows for the advantages of efficiency to be realized, thus providing much needed services at lower cost, and relieving the burden of overcrowding in traditional medical practices. These advantages are particularly relevant to simple preventive services that are needed by a large fraction of the population on a frequent basis. 

We analyzed preventive services to identify the organizational structure that would be most effective in delivering them and determine whether they would benefit from high-efficiency processes that are characteristic of retail service organizations.  For each service we evaluated the number of delivery repetitions that are required for the target population; the time, number of personnel and cost required to deliver the services in physician-based practices; the level of training required for performance of preventive services; and what reductions in time and cost would be obtained through efficiency in a retail context. 

\section{Methods}
\label{sec:methods}

We estimated the required number of repetitions (from target population and frequency), time and cost in the traditional healthcare setting for 35 of the services recommended by United States Preventive Services Task Force (USPSTF) [4], the Advisory Committee on Immunization Practices (ACIP) [17], and the Centers for Medicare and Medicaid Services (CMS) [18]. Where recommendations specify ``regularly,'' e.g., for short informational counseling sessions, we assume annual frequency. Medication prescriptions were incorporated into the frequency of the related contact service. The time to provide the service includes the contact and related administrative time estimated using 2007 CMS Physician Fee Schedule (PFS) [19] in Relative Value Units (RVUs), which can be converted consistently to time based on one RVU as equivalent to 0.5 hours of physician time, as stated in military policy and studies (15.4 RVUs for 8 hours in 2003) [20]. For cases where recommended and estimated contact time are both available, we found them to be similar, with one exception: diabetes self-management is recommended to be 30 contact minutes, while the RVU based estimate is 22.5 minutes. For this case, we use the recommended contact time. 

Unless otherwise specified, costs are from CMS specifications [19, 21, 22]. We did not include the allocation for malpractice insurance, which for each service is under 5\% of total costs, or the ``budget neutrality factor.'' Varying costs are those expected to decline when services are subject to process streamlining. This includes all contact and administrative times. Fixed costs include chemoprophylaxis medications and immunization vaccines because medication manufacture and delivery is already a combination of patent protection and industrial efficiency. Multiple service options or medication types were averaged by usage and cost data [22, 23]. Laboratory costs were treated as varying because rates [21] do not use existing lower cost options (e.g., over-the-counter immunologic fecal occult blood tests) or have remained fixed over time (e.g., laboratory rates for serum potassium and creatinine, 2002--2007). Current total cost estimates use available coverage data [7]. Where necessary, costs were inflated to 2007 using 3\% annual increases.

We evaluated the potential efficiency gains from process improvement including shorter times for task execution and lower costs. The empirical ``learning curve'' [24--26] quantifies increasing efficiency for tasks performed repetitively by individuals and for entire industries. The relationship between the time $t_1$ required to perform the first repetition and the time required for the $n$th repetition, $t_n$, is
\begin{displaymath}
t_n = t_1 n^{-z}
\end{displaymath}
The learning-curve parameter $z$ varies between industries, clustering around 0.15 and generally in the range (0.07, 0.4). We used $z = 0.15$ for our estimates and conservatively considered only individual provider repetitions, not industry-wide repetitions, to estimate efficiency gains. The conclusions are robust to varying these assumptions. We did not directly consider specific mechanisms for efficiency, e.g., group informational sessions. Costs follow a similar behavior. However, we considered the base (first repetition) cost to be half the physician provider costs, because lower-salaried employees can provide high-repetition, low-complexity services. 

The ``learning curve'' neglects idle time---the true cost and time for a retail employee depends on the demand. Retail demand depends on the level of promotion by retailers themselves. We made calculations for an individual provider that performs $m = 20$ distinct services; increasing $m$ reduces the efficiency gains, because it reduces the number of repetitions of each service.  However, an increased number of services increases the likelihood of sufficient demand to utilize employee time. A change of $m$ by 10\% would change the efficiency gain by 2\%. The efficiency gains are affected by time estimates through the number of repetitions possible for each provider per year. Where time is estimated roughly due to lack of direct specification (e.g., short counselling sessions) the efficiency factors are robust as even an increase of 100\% or decrease of 50\% (a factor of two) yields a change in efficiency of 12\%. 

To illustrate the effect of efficiency gains, we compared costs and personnel requirements for full coverage by current providers with retail organizations in the third year of operation (from level budgeting in each of three years). The number of services provided by an individual was obtained using 1824 hours per year [27]. Our RVU-derived times underestimate personnel due to the use of physician cost equivalents for administrative time. For tobacco cessation we also calculated the impact of quits on the target population reducing the costs and personnel. We developed a simple model consistent with available data [28] using an exponentially decreasing quit rate saturating at a total of 23\% quits. 

We also estimate the reduction of acute care costs due to full preventive coverage based upon reviews of the National Commission on Prevention Priorities, which take into account efficacy and adherence [29]. Tobacco cessation healthcare cost reductions were estimated by an exponential decay with a conservative three-year time constant starting in the year following a quit.

We also evaluated which services would be appropriate based upon their apparent complexity, reflected in protocols, diversity of actions, or required certification. Explicit protocols for services and levels of training for providers are available in the relevant literature.

We identified other features of physician offices, retail settings, and dedicated healthcare providers that would be important for the customer or for the physician's ability to provide preventive services to the target population. These include population served, location relative to target population, and core competencies.

\section{Results}
\label{sec:results}

The number of repetitions needed for full coverage of the target population, the time to deliver the service, and the variable and fixed costs are summarized in Table 1 (Appendix). Across the analyzed services, the number of repetitions ranges from 380,000 abdominal aortic aneurysm ultrasounds to 560 million obesity-information counseling sessions. Repetitions over 100 million also include information sessions for tobacco cessation, hyperlipidemia diet, problem drinking, aspirin chemoprophylaxis, calcium chemoprophylaxis, diabetes nutrition and screening for hypertension and cholesterol.  We excluded depression screening because recommendations state [7] that this service should be provided by a facility that can also provide treatment.

Very few of the preventive services require particular levels of training. The risks associated with preventive services are generally much lower than those of acute care. \mbox{Mammography} is distinct because high levels of training are required to interpret the images, but not to acquire them; the latter is currently performed in some retail settings [30]. There are also tasks that require ongoing organizational mechanisms, such as the timing and record keeping of childhood immunizations, but these were not considered to preclude retail delivery.

Consider tobacco cessation information sessions: The USPSTF [7] recommends one session per year, but only 28\% of smokers [10] receive such sessions even though over 70\% of smokers indicate they want to quit [31]. CMS authorizes up to eight sessions per year [18]. The benefits have been studied [28] extensively: The estimated quit rate from a single session is 2.5\%, 5\% with cessation aids. Multiple sessions raise the quit rate to about 23\%. On average, each smoker who quits gains about six years of life expectancy and reduces per annum adult healthcare costs by \$1,100 (from \$10,200 to \$9,100, i.e. 12.5\%), with most of the benefits accruing after 3 years. An information session has been specified by a simple protocol [32] composed of questions such as: ``Does patient now use tobacco?  Is patient now willing to quit? \ldots''

Our estimates suggest that providing tobacco cessation information through retail organizations would reduce the cost and effort of providing the service. The per-person cost of the session (\$11.40) and medications (\$209 per year for the 16\% who use them) is \$34.  If the standards were increased universally to eight times per year (as suggested by CMS)---370 million sessions annually---the cost in the current system would be \$6.2 billion and would require 34,000 full-time personnel for that service alone. By contrast, full coverage may be achieved in a retail setting in the third year at a cost of \$1.7 billion per year and 7,900 personnel (if we do not include the reduction of the target population due to quits, the values would be \$2.25 billion and 9,700 personnel). By the end of the third year 21\% of smokers would be expected to quit, constituting 30\% of the individuals who say they want to quit, and 4\% of the population as a whole. We estimate \$1.57, \$3.77 and \$5.83 billion in acute care savings in years 2, 3 and 4 respectively.  Differences in healthcare costs and life expectancy [33, 34] imply these quits eventually save \$8.9 billion per year in acute care costs (\$9.7 billion for a budget of \$2.25 billion per year due to improved coverage in the first two years), and result in a combined life expectancy gain of 59 million years.

Contrast the case of colonoscopies for colorectal cancer screening: USPSTF recommends one per ten years [7, 35] with alternatives of sigmoidoscopy or fecal occult blood tests. The number of colonoscopies required is relatively small (low scale), only 2 million per year. The procedure itself and the common use of anesthesia are significantly more complex than tobacco cessation counseling. Thus, high service complexity and low scale make colonoscopies not well suited for a retail setting.  

Our calculations show that we can achieve a five-fold reduction in variable costs of simple large-scale services if they are delivered by efficiency-oriented organizations (Table 1). Such efficiency gains can be achieved with 500 repetitions of a service per provider per year (a factor of 2.5 in time reduction and factor of 2 reduction in base costs of employees). Where medication costs are a significant fraction of total costs, efficiency gains for the total service are smaller.

A significant barrier to full implementation of clinical preventive services in the existing healthcare system is the actual workload burden. While recognized in publications [36] this realization has yet to influence policy, perhaps due to a lack of identified alternatives. We estimate full delivery in the current system would require over 400,000 full-time personnel for these services alone (Figure 1). A retail context would reduce the required number of personnel through efficiency gains, reduce the level of training required, and place the services in an industry that can more readily deploy the necessary workforce.

\begin{figure}[tb]
\includegraphics{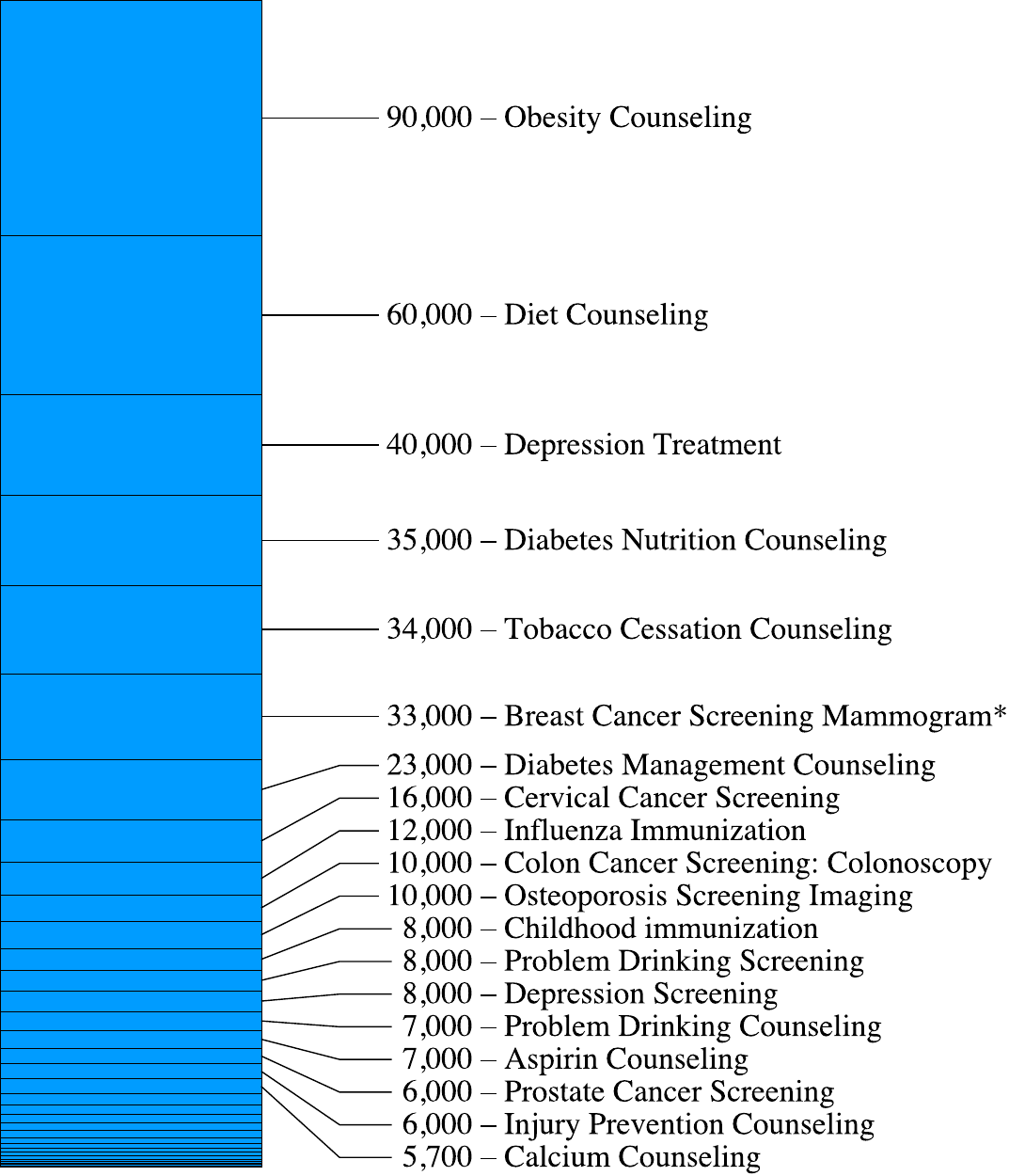}
\caption{\label{fig:bar} Personnel requirements for fully delivering clinical preventive services in the current healthcare provider setting. *Mammography includes both acquisition and interpretation.}
\end{figure}

\section{Discussion}
\label{sec:discussion}

Performing repetitive tasks using a suitable process results in dramatic efficiency gains. While such an approach does not apply to most medical care, it can apply to flu shots and other simple tasks. Historically, industrial efficiency gains resulted in widespread availability of key goods and services that were not available previously. The industrial revolution was in large part driven by mass production. Model-T Ford cars were introduced in 1909 at \$850, a significantly lower price than competitors and further declined in price to \$260 in the 1920s [37, 38], consistent with the empirical ``learning curve'' [24, 25]. The time required to assemble a single car declined from 12.5 hours in earlier production systems to 93 minutes [37, 38]. The total production of Model-Ts was 15 million, a fraction of the annual need for many preventive services today. While the Model-T production line example is not directly applicable to health services, the approach of high efficiency is today similarly embedded in retail services. 

Our analysis suggests that widespread retail availability could dramatically improve the delivery of preventive services to underserved populations in the US. 

We caution that misapplication of efficiency to complex tasks that require careful decision making by highly-trained individuals would lead to ineffective execution. Because individual actions are distinct, repetition should not be expected to lead to efficiency gains. With continued pressure to reduce healthcare costs, it is imperative to distinguish those services to which efficiency can be applied and those to which it cannot. Without such differentiation, efforts to reduce costs would lead to less effective services, rather than increased efficiency.

The largest-scale preventive services are most appropriate for retail organizations. Setting the threshold at services corresponding to 500 per 10,000 population (or 250 for each of the 59,000 US pharmacies [39]) and avoiding complex services, we obtained the set marked as retail-appropriate in Figure 1. Lifestyle and chemoprophylaxis counseling is the largest set of high benefit services for retail implementation.

The total cost of the identified retail-appropriate services that would be transferred from physician to retail delivery by this model, excluding costs that are not transferred, such as medications, are approximately \$14 billion (allowing delivery of services for which no estimates are available to be at 20\%), corresponding to 3.2\% of national expenditures on physician and clinical services [40]. Why should physicians endorse the transfer of healthcare tasks to others, even such a small percentage? Because unlike other services, physicians are not necessary to guarantee quality delivery; because of the resulting dramatic increase in delivery of needed prevention services; because of the reduction of pressures to deliver these services; to enable physicians to focus on tasks requiring their higher training; and because improving the efficiency of the healthcare system allows more resources to be devoted to tasks that need them, and increases the ability of the system to meet demands beyond the current capacity of providers and resources. Everyone, including traditional providers, would benefit from improved healthcare system function, especially if reimbursement is made appropriate by recognizing the distinct needs for simple and complex tasks.

Using a retail context for a restricted set of services is consistent with the existing US healthcare system. Prescription pharmacies provide a service for which physician offices are no longer considered practical. We offer four factors that characterize why physician offices do not (and should not be expected to) provide highly efficient services, including preventive services: The population served, the location of service (travel distance), the number of locations at which the service can be obtained, and the frequency at which the target population is present at the location of service for other reasons. The population served by a physician office is limited. Efficiency arises from repetition and with fewer repetitions, the efficiency gain is smaller. Moreover, efficiency gains in a practice setting cannot be translated into more individuals served. The exclusive relationship between the patient and the physician precludes patient choice. Thus, the drivers for efficiency improvement and cost reduction to compete for patients are not present. Most importantly, the physician office is designed for high-complexity tasks. Replacing this focus to provide efficient delivery would reduce the effectiveness of its primary responsibility.

For preventive services, a retail concept is particularly convenient for healthy individuals, since multiple locations can serve any individual as opposed to the exclusive physician model. This results in lower travel and waiting times. The importance of sufficient demand motivates retailers to promote their preventive services, which would benefit population health. The non-urgent nature of these services reduces their priority among both traditional providers and patients. Wrapping preventive care in the nationally popular focus on health and fitness, rather than in a medical context, may provide new opportunities for wider adoption. Given the empirical evidence that the existing US healthcare system cannot improve upon its current performance efforts, delivering preventive services through retail operations should be supported by policy.

\section{Acknowledgements}
\label{sec:acknowledgements}

This work was supported in part by the Centers for Disease Control and Prevention.

\newpage
\section{Appendix}

\begin{figure}[h!]
\centerline{\includegraphics[scale=0.9]{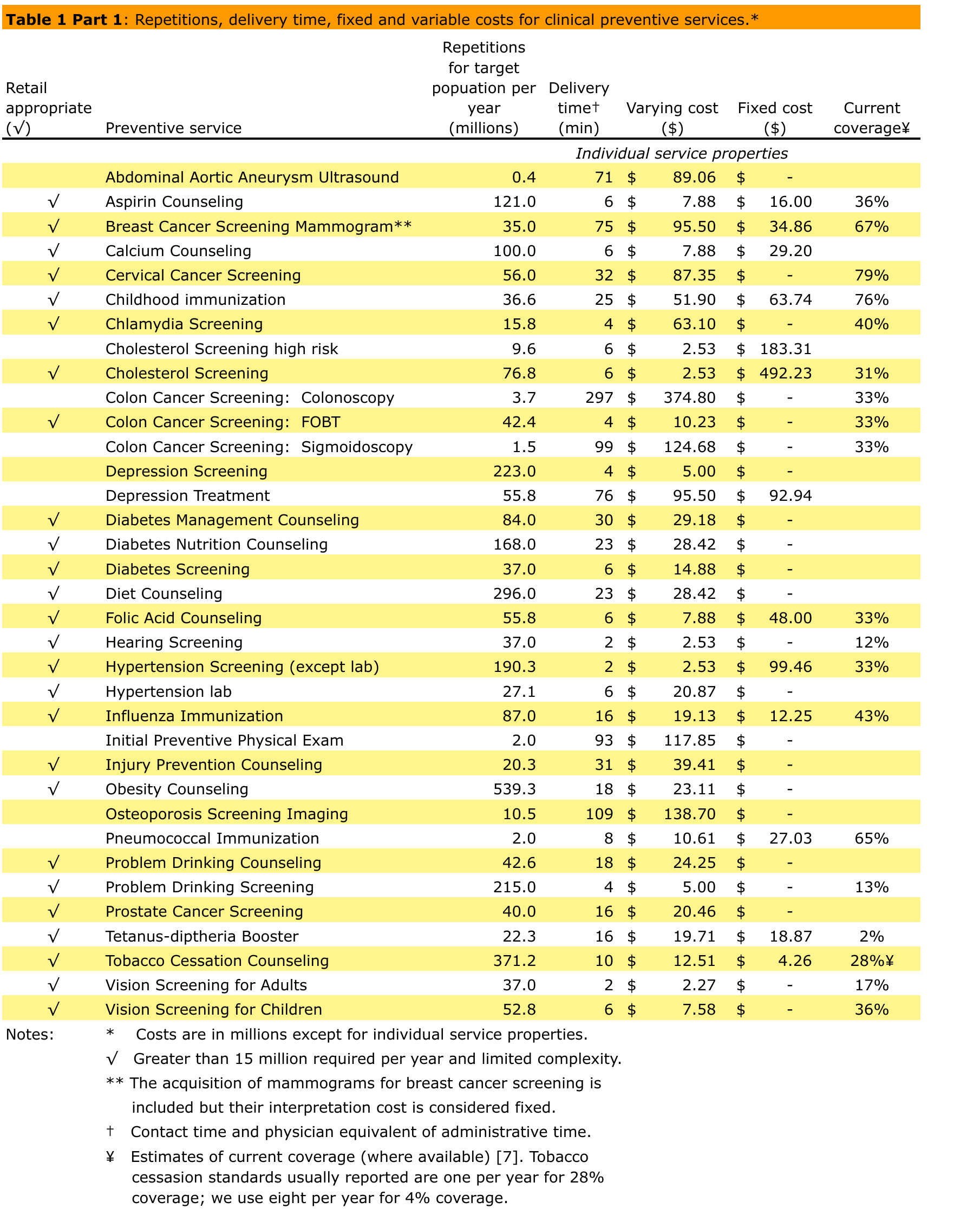}}
Table 1 Part 1: Key properties for retail implementation of preventive services.
\end{figure}

\begin{figure}[t!]
\centerline{\includegraphics[scale=0.90]{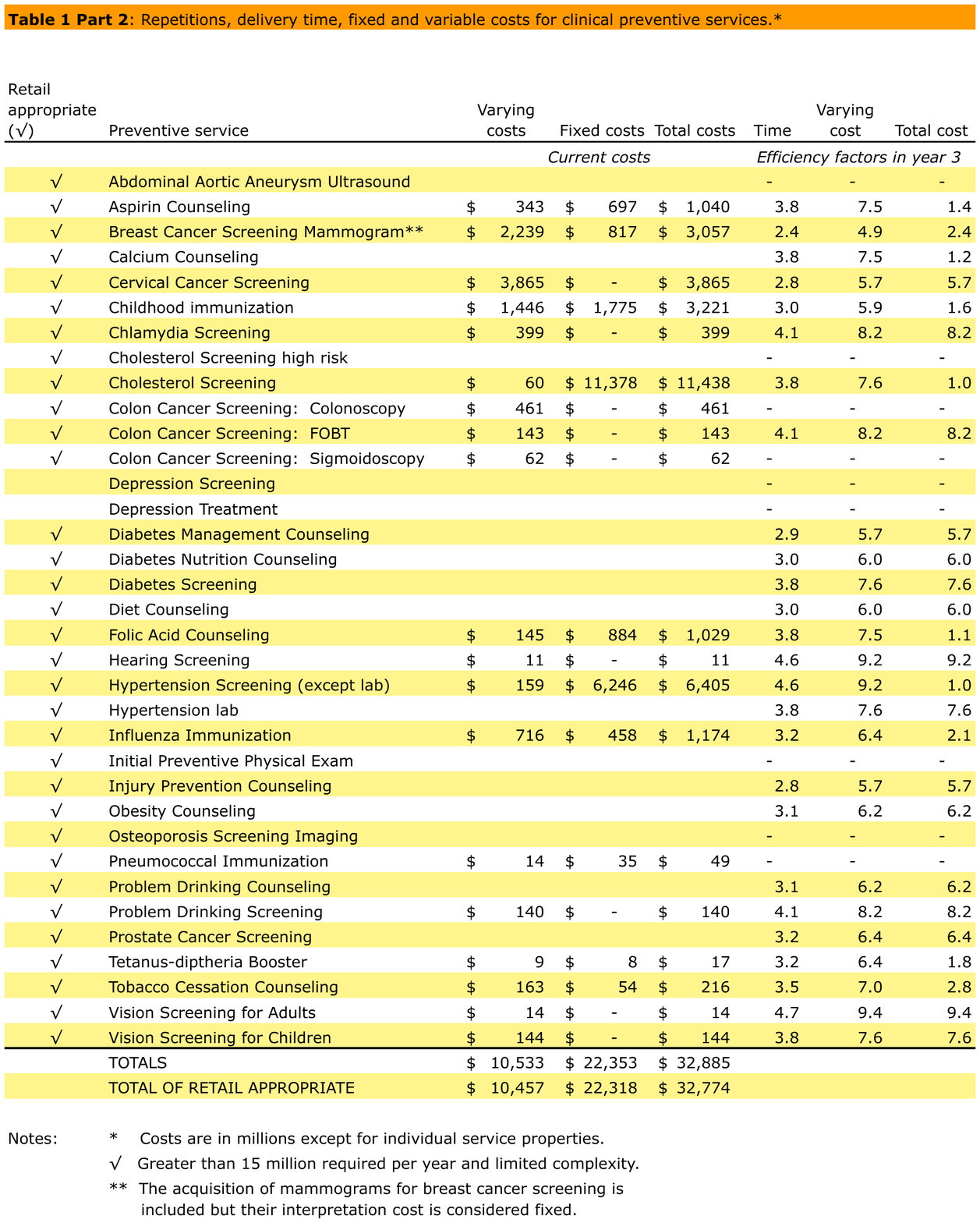}}
Table 1 Part 2: Key properties for retail implementation of preventive services.
\end{figure}

\begin{figure}[t!]
\centerline{\includegraphics[scale=.95]{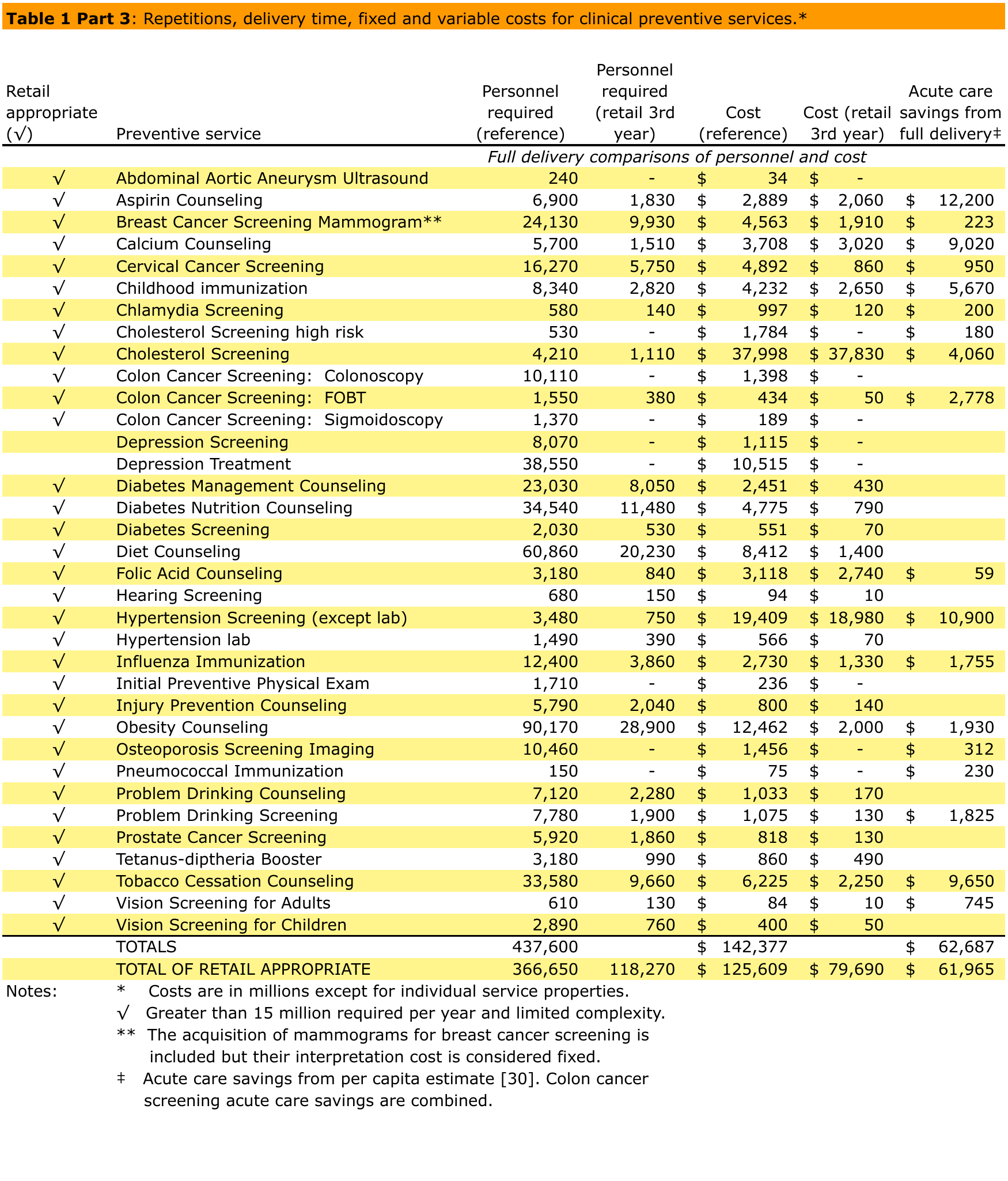}}
Table 1 Part 3: Key properties for retail implementation of preventive services.
\end{figure}

\clearpage
\newpage

\section{References}

\begin{enumerate}
\item Institute of Medicine, Committee on Quality of Health Care in America. Crossing the quality chasm: a new health system for the 21st century (National Academy Press, Washington DC, 2001).
\item  J. Lambrew, Who has the cure?: Hamilton Project ideas on health care (Brookings Institution Press, Washington, DC, 2008), p227.
\item  Y. Bar-\!Yam, Improving the effectiveness of health care and public health: A multi-scale complex systems analysis, {\it American Journal of Public Health} {\bf 96}, 459--66 (2006).
\item  Y. Bar-\!Yam, Making things work (Knowledge Press, Cambridge, MA, 2005). Available from: \url{http://necsi.edu/publications/mtw/}.
\item  Y. Bar-\!Yam, Multiscale variety in complex systems, {\it Complexity} {\bf 9}, 37--45 (2004). Available from: \url{http://necsi.edu/research/multiscale/}.
\item  Centers for Disease Control and Prevention, Smallpox: 30th anniversary of global eradication. (Centers for Disease Control and Prevention, Atlanta, GA, 2008). Available from: \url{http://www.cdc.gov/Features/SmallpoxEradication}.
\item  Agency for Healthcare Research and Quality, U.S. Preventive Services Task Force guide to clinical preventive services, Report number: 07-05100 (U.S. Department of Health and Human Services, 2007).
\item  M. Maciosek, A. Coffield, N. Edwards, T. Flottemesch, M. Goodman, L. Solberg, Priorities among effective clinical preventive services: results of a systematic review and analysis, {\it American Journal of Preventive Medicine} {\bf 31}, 52--61 (2006).
\item  American Medical Association, Promoting healthy lifestyles (2008). Available from:  \url{http://www.ama-assn.org/ama/pub/physician-resources/public-health/promoting-healthy-lifestyles.page}.
\item  National Commission on Prevention Priorities, Preventive care: a national profile on use, disparities, and health benefits (Partnership for Prevention, Washington DC, 2007).
\item  Agency for Healthcare Research and Quality, Guide to clinical preventive services. 2nd ed. (U.S. Department of Health and Human Services, Rockville, MD, 1996).
\item  M. K. Scott, Health care in the express lane: the emergence of retail clinics (California Healthcare Foundation, Oakland, CA, 2006). Available from: \url{http://www.chcf.org/topics/view.cfm?itemID=123218}.
\item  R. Bohmer, The rise of in-store clinics---threat or opportunity? {\it New England Journal of Medicine} {\bf 356}, 765--68 (2007). 
\item  M. B. Marcus, Medical clinics in retail settings are booming. {\it USA Today} (28 Aug 2011). Available from: \url{http://yourlife.usatoday.com/health/healthcare/story/2011-08-28/Medical-clinics-in-retail-settings-are-booming/50168796/1}
\item  Deloitte Center for Health Solutions, 2008 Survey of health care consumers. (Deloitte Development LLC, Washington DC, 2008).
\item American Medical Association, AMA calls for investigation of store-based health clinics (2007). Available from: \url{http://www.medicalnewstoday.com/articles/75308.php}. 
\item  CDC Advisory Committee on Immunization Practices, Immunization schedules (Centers for Disease Control and Prevention, Atlanta, GA, 2008). Available from: \url{http://www.cdc.gov/vaccines/recs/schedules}. 
\item  Centers for Medicare \& Medicaid Services, Medicare preventive services quick reference information (2008). Available from: \url{http://www.cms.hhs.gov/MLNProducts/downloads/MPS_QuickReferenceChart_1.pdf}. 
\item  Centers for Medicare \& Medicaid Services. Physician fee schedule relative value files (2007). Available from: \url{http://www.cms.hhs.gov/PhysicianFeeSched/}. 
\item  J. St. Andrews, A study of the Relative Value Unit as a practice management tool for provider productivity (Army Medical Department, Miami, FL, 2003). Available from: \url{http://www.dtic.mil/cgi-bin/GetTRDocLocation=U2&doc=GetTRDoc.pdf&AD=ADA421282}. 
\item  Centers for Medicare \& Medicaid Services, Clinical laboratory fee schedule (2007). Available from: \url{http://www.cms.hhs.gov/ClinicalLabFeeSched/}. 
\item  Centers for Medicare \& Medicaid Services, Medicare Part B drug average sales price (2007). Available from: \url{http://www.cms.hhs.gov/McrPartBDrugAvgSalesPrice/}. 
\item  Verispan VONA, Top 200 drugs in 2006 (2006). Available from: \url{http://drugtopics.modernmedicine.com/drugtopics/data/articlestandard//drugtopics/092007/407652/article.pdf}. 
\item  B. Henderson, The experience curve reviewed, parts I through IV. {\it Perspectives}, {\bf 124} (1973).
\item L. E. Yelle, The learning curve: historical review and comprehensive survey. {\it Decision Science} {\bf 10}, 302--28 (1979). 
\item  A. Newell, P. S. Rosenbloom, Mechanisms of skill acquisition and the law of practice. In: J. R. Anderson (ed), Cognitive skills and their acquisition (Lawrence Erlbaum Associates, Hillsdale, NJ, 1981). p1--51.
\item  L. Mishel, J. Bernstein, S. Allegretto, The state of working in America 2006/2007 (ILR Press, Ithaca, NY, 2007).
\item  L. J. Solberg, M. V. Maciosek, N. M. Edwards, H. S. Khanchandani, A. L. Butani, D. A. Rickey, M. J. Goodman, Tobacco use screening and counseling: technical report prepared for the National Commission on Prevention Priorities (HealthPartners Research Foundation, Minneapolis, MN, 2006).
\item  Partnership for Prevention, Rankings of preventive services for the U.S. population (2008). Available from: \url{http://www.prevent.org/National-Commission-on-Prevention-Priorities/Rankings-of-Preventive-Services-for-the-US-Population.aspx}. 
\item  See e.g. Baptist Memorial Health Care and Goldsmith's open Tennessee's first retail-based mammography center (2008). Available from: \url{http://www.bmhcc.org/media/news/archivecontent.asp?category=2000&article=News_Item.2004-11-12.7860892798}.
\item  Centers for Disease Control and Prevention, Cigarette smoking among adults---US 2004. {\it Morbidity and Mortality  Weekly Report} {\bf 54}, 44 (2005). 
\item  U.S. Department of Health \& Human Services, Screen for tobacco use status (2007). Available from: \url{http://www.surgeongeneral.gov/tobacco/screen.htm}.
\item  Centers for Disease Control and Prevention, Annual smoking-attributable mortality, years of potential life lost, and productivity losses---U.S., 1997-2001. {\it Morbidity and Mortality Weekly Report} {\bf 54}, 625--28 (2005).
\item  Centers for Disease Control and Prevention, Cigarette smoking attributable morbidity---U.S., 2000. {\it Morbidity and Mortality Weekly Report} {\bf 52}, 842--44 (2003).
\item  M. V. Maciosek, L. J. Solberg, A. B. Coffield, N. M. Edwards, H. S. Khanchandani, A. L. Butani, D. A. McGree, M. J. Goodman, Colorectal cancer screening: technical report prepared for the National Commission on Prevention Priorities (Partnership for Prevention, 2006). Available from: \url{http://www.prevent.org/data/files/initiatives/colorectalcancerscreening.pdf}. 
\item  K. Yarnall, K. Pollak, T. Ostbye, K. Krause, J. Michener, Primary care: is there enough time for prevention? {\it American Journal of Public Health} {\bf 93}, 635--41 (2003). 
\item  Model T Facts (Ford Motor Company, Detroit, MI, 2008). Available from: \url{http://media.ford.com/article_display.cfm?article_id=858}. 
\item  Assembly line (Wikipedia, 2001). Available from: \url{http://en.wikipedia.org/wiki/Assembly_line}. 
\item  Pharmaceutical Care Management Association, Consumer access to pharmacies in the United States (2007). Available from: \url{http://pcma-upgrade.webdrivenhq.com/assets/2008-03-25_Research_SKA\%20Research\%20Consumer\%20Access\%20to\%20Pharmacies\%202007\%205_1_07.pdf}. 
\item  Dept. of Health \& Human Services, National health expenditure accounts, 2006 highlights (2008). Available from \url{http://www.cms.hhs.gov/NationalHealthExpendData/downloads/highlights.pdf}.
\end{enumerate}

\end{document}